\documentclass[preprint,aps,tightenlines,showpacs]{revtex4}

\usepackage[dvips]{graphicx}
\usepackage{dcolumn}
\usepackage{bm}

\newcommand{\sla}[1]{\not\! #1}
\def\pnus{|\bm{p}_\nu|}
\def\ples{|\bm{p}_l|}

\def\pqs{|\bm{q}|}
\def\ppisc{|\bm{k}_c|}
\def\pqsc{|\bm{q}_c|}

\def\cphi{\cos\phi_\pi}
\def\sphi{\sin\phi_\pi}
\def\cphit{\cos 2\phi_\pi}
\def\sphit{\sin 2\phi_\pi}
\def\sh{\sinh \alpha}
\def\ch{\cosh \alpha}

\begin{document}
 
{  

\title{Dynamical Model of Weak Pion Production Reactions}

\author
{ Toru Sato$^{1,}$\cite{ts},
  D. Uno$^{1,}$\cite{du},
  T.-S. H. Lee$^{2,}$\cite{tshl} }

\affiliation
{$^1$ Department of Physics, Osaka University, Toyonaka, Osaka 560-0043, Japan \\
$^2$Physics Division, Argonne National Laboratory, Argonne, Illinois 60439}

\date{\today}

\begin{abstract}

The dynamical model of pion electroproduction developed in Physical
Review C63, 055201 (2001) has been extended to investigate
the weak pion production reactions. With the Conserved Vector Current(CVC)
hypothesis, the weak vector currents are constructed from electromagnetic
currents by isospin rotations. Guided by the effective chiral lagrangian method
and using the unitary transformation method developed previously, the weak
axial vector currents for $\pi$ production
are constructed with no adjustable parameters.
In particular, the N-$\Delta$
transitions at $Q^2\rightarrow 0$ 
are calculated from the constituent quark model
and their $Q^2$-dependence is assumed to be identical to that determined
in the study of pion electroproduction. The main feature of our approach
is to renormalize these bare N-$\Delta$ form factors with the
dynamical pion cloud effects originating from the non-resonant $\pi$ production
mechanisms.  The predicted cross sections
of neutrino-induced pion production reactions, $N(\nu_\mu, \mu^- \pi)N$,
 are in good agreement with the existing data. We show that the 
renormalized(dressed) axial N-$\Delta$ form factor contains
large dynamical pion cloud effects and this renormalization effects are
crucial in getting agreement with the data. We conclude that the N-$\Delta$
transitions predicted by the constituent quark model are consistent with 
the existing neutrino induced
pion production data in the $\Delta$ region, contrary to the previous
observations. This is consistent with  our previous findings in the
study of pion electroproduction reactions. However, more extensive and
precise data of neutrino induced pion production reactions
are needed to further test our model and to pin down the
$Q^2$-dependence of the axial vector N-$\Delta$ transition form factor.
\end{abstract}

\maketitle
}
\section{INTRODUCTION}

It is now well recognized that 
an important challenge in
nuclear research is to understand the hadron structure 
 within Quantum Chromodynamics(QCD). 
One of the important information for pursuing this research is 
the electromagnetic and weak
 N-$\Delta$ transition form factors. Such information are 
important for testing current hadron models and perhaps also
Lattice QCD calculations in the near future. In particular, we can address
the question about whether N and $\Delta$ are deformed. 
If they are deformed, is it due to the gluon interactions between
quarks or due to the pion cloud which is resulted from the spontaneously
breaking of the chiral symmetry of QCD?

As a step in this direction, 
we have developed\cite{sl1,sl2} in recent years a dynamical model for
investigating the pion photoproduction and
electroproduction reactions at energy 
near the $\Delta$ resonance. The dynamical approach, which has also been
pursued by several 
groups\cite{tanabe,yang,nbl,pearcelee,afn,surya,ky1,ky2},
 is different from the
approaches based on either dispersion relations\cite{cgln,hdt,a98} or
 K-matrix method\cite{muko90,muko99,maid,said} in interpreting the data.
In Refs.1 and 2, we not only have extracted the electromagnetic 
N-$\Delta$ transition form factors
from the data, but
also have provided an interpretation of the extracted
parameters in terms of hadron model calculations.
In particular, we have shown that including the dynamical
pion cloud effect in theoretical analyses
can resolve a long-standing puzzle that  
the N-$\Delta$ M1 transition form 
factor $G_M^{N\Delta}(0)$ predicted by the
constituent quark model is about 40 $\%$ lower than the empirical value.
Furthermore, the dynamical pion cloud is found to play an important role
in determining the $Q^2$-dependence of 
$G_M^{N\Delta}(Q^2)$ which drops faster than the
proton form factor $G_p(Q^2)$ 
as $Q^2$ increases. The predicted electric E2 and Coulomb C2 
N-$\Delta$ transition form factors,
which exhibit very pronounced meson cloud effects at low $Q^2$, have
stimulated some experimental initiatives\cite{bern02}. Similar results 
were also obtained in Ref.\cite{ky2}.

Since weak currents are closely related to electromagnetic currents
within the Standard Model, it is straightforward  to extend our 
dynamical approach to study weak pion production reactions.
In this paper, we  report the progress we have made in this 
direction.
Our objective is to develop a framework for extracting 
the  axial  vector N-$\Delta$ form factors from the
data of neutrino-induced pion production reactions.  
In particular, we would like to explore whether a dynamical approach
, as developed in our study of pion electroproduction, can
also resolve a similar problem that
 the axial N-$\Delta$ transition strength calculated\cite{hhm,lmz}
from the constituent quark model is
about 30 $\%$ lower than what was extracted\cite{data90} from the data.
We will also make predictions for future experimental
tests which could be conducted at Fermi Lab, KEK and Japan
Hadron Facility  in the near future.

To introduce our model, it is necessary to first briefly review
the dynamical approach developed in Ref.\cite{sl1,sl2}
(called SL model). Our starting point is an
interaction Hamiltonian 
$H_{I} = \Gamma_{MB\leftrightarrow B'}$ which describes
the absorption and emission of mesons($M$) by  baryons($B$).
In the SL model, such a Hamiltonian is obtained 
from  phenomenological Lagrangians.
In fact, the approach is more general and this Hamiltonian can be
defined in terms of a hadron model, as attempted, for example, 
in Ref. \cite{yoshi}, or by using the vertex functions predicted
by Lattice QCD calculations.  

It is a non-trivial many-body problem to 
calculate $\pi N$ scattering and $\gamma N \rightarrow \pi N$ 
reaction amplitudes from the interaction Hamiltonian
$H_{I} = \Gamma_{MB\leftrightarrow B'}$. 
To obtain a manageable reaction model,
a unitary transformation method\cite{sl1,ksh} is used  
up to second order in $H_I$ to derive an effective Hamiltonian.
 The essential idea of the employed
unitary transformation method is to eliminate the unphysical vertex
interactions $MB \rightarrow B^\prime$ with $m_M +m_B < m_{B^\prime}$ 
from the Hamiltonian and absorb their effects into 
 $MB\rightarrow M^\prime B^\prime$ two-body interactions. 
For the pion production processes in the $\Delta$ region,
the resulting effective Hamiltonian is  
defined in a subspace spanned by the
 $\pi N$, $\gamma N$ and $\Delta$ states and has the following form
\begin{eqnarray}
H_{eff} & = & H_0 + v_{\pi N} + v_{\gamma\pi}
 + \Gamma_{\pi N \leftrightarrow \Delta} 
+ \Gamma_{\gamma N \leftrightarrow \Delta},   \label{hamile}
\end{eqnarray}
where
$v_{\pi N}$ is a non-resonant $\pi N$ potential, and $v_{\gamma\pi}$
describes the non-resonant $\gamma N \leftrightarrow \pi N$
transition. The $\Delta$ excitation
is described by the vertex interactions
 $\Gamma_{\gamma N \leftrightarrow \Delta}$
 for  the $\gamma N \leftrightarrow\Delta$ transition
 and $\Gamma_{\pi N \leftrightarrow \Delta}$ 
for the  $\pi N \leftrightarrow\Delta$ transition. 
The non-resonant $v_{\gamma\pi}$ consists of the usual pseudo-vector
Born terms, $\rho$ and $\omega$ exchanges, and the crossed $\Delta$
term. 

From the effective Hamiltonian Eq.~(\ref{hamile}), it is straightforward 
to derive
a set of coupled equations for $\pi N$ and $\gamma N$ reactions.
The resulting pion photoproduction amplitude can be written as
\begin{eqnarray}
T_{\gamma\pi}(E)  =  t_{\gamma\pi}(E) + 
\frac{
\bar{\Gamma}_{\Delta \rightarrow \pi N}(E)
\bar{\Gamma}_{\gamma N \rightarrow \Delta}(E)
}
{E - m_\Delta - \Sigma_\Delta(E)} . \label{tmatt}
\end{eqnarray}
The non-resonant amplitude $t_{\gamma\pi}$ is 
calculated from $v_{\gamma\pi}$
by
\begin{eqnarray}
t_{\gamma\pi}(E)=   v_{\gamma\pi} 
+  t_{\pi N}(E)G_{\pi N}(E)v_{\gamma\pi},  \label{tmatg}
\end{eqnarray}
where $G_{\pi N}$ is the $\pi N$ free propagator, 
and $t_{\pi N}$ is calculated from 
the non-resonant $\pi N$ interaction $v_{\pi N}$.

The dressed vertices in Eq.~(\ref{tmatt}) are defined by 
\begin{eqnarray}
\bar{\Gamma}_{\gamma N \rightarrow \Delta}(E)  &=&  
   \Gamma_{\gamma N \rightarrow \Delta} + v_{\gamma \pi} G_{\pi N}(E)
\bar{\Gamma}_{\pi N \rightarrow \Delta}(E)  ,  \label{vertg} \\
\bar{\Gamma}_{\Delta \rightarrow \pi N}(E)
 &=& [1+t_{\pi N}(E)G_{\pi N}(E)]\Gamma_{\Delta\rightarrow\pi N}. \label{vertp}
\end{eqnarray}
The $\Delta$ self-energy in Eq.~(\ref{tmatt}) is then calculated from
\begin{eqnarray}
\Sigma_\Delta(E) = 
 \Gamma_{\pi N\rightarrow \Delta}
G_{\pi N}(E)\bar{\Gamma}_{\Delta \rightarrow \pi N}(E).  \label{self}   
\end{eqnarray}

As seen in the above equations,
an important feature of the dynamical model is that 
the bare vertex $\Gamma_{\gamma N\rightarrow \Delta}$ is modified 
by the non-resonant interaction $v_{\gamma\pi}$ to give the dressed vertex 
$\bar{\Gamma}_{\gamma N\rightarrow\Delta}$, as defined by Eq.~(\ref{vertg}). 
The second terms in the right-hand sides of Eqs.~(4) and (5) can be 
interpreted as the 
$dynamical$ pion cloud effects on the N-$\Delta$ transitions.
We can then identify the parameters of the bare vertex 
$\Gamma_{\gamma N \rightarrow \Delta}$ with the predictions from
hadron models within which the $\pi$-baryon states in $continuum$
are excluded.
Such a separation of the reaction mechanisms from the excitations of
hadron internal structure is an essential ingredient of a dynamical approach,
but is not an objective of the approaches
based on either dispersion relations\cite{cgln,hdt,a98} 
or K-matrix method\cite{muko90,muko99,maid,said}.

In the above formulation, the matrix elements $v_{\gamma \pi}$
and $\Gamma_{\gamma N \leftrightarrow \Delta}$ are determined by the
electromagnetic current $J^{em}_\mu$.
The dynamical formulation for investigating weak pion production
 reactions can be obtained from Eqs.~(1)-(6) by replacing $J^{em}_\mu$ by 
$V_{\mu}-A_{\mu}$, where $V_\mu$ and $A_\mu$ are the weak vector and
axial vector currents respectively.
By Conserved Vector Current(CVC) relations,
the vector currents $V_\mu$ can be
obtained from the electromagnetic currents by isospin rotations. 
Guided by the effective chiral Lagrangian method and
 using the similar procedures of the
SL model, we can construct the axial vector
currents $A_\mu$ for pion production reactions. 
The resulting axial current
amplitude consists of  nucleon-Born, rho-exchange, pion-pole,
and $\Delta$ excitation terms and can be calculated using the
parameters in the literatures. 
In particular, the bare axial N-$\Delta$ transitions at $Q^2\rightarrow 0$ 
are assumed to be those calculated by Hemmert, Holstein, and 
Mukhopadhyay\cite{hhm} within the constituent quark model.

Most of the earlier theoretical
investigations\cite{salin,adler,bij,zuker,schr73} of
weak pion production reactions were performed
during the years around 1970.
The model which was most often used in analyzing the data 
was developed by Adler\cite{adler}. He considered a model based on
the dispersion relations of Chew, Goldberger, Low, and Nambu(CGLN)
\cite{cgln}. The driving terms and subtraction terms of the
dispersion relations
are calculated from the nucleon Born terms. 
No vector meson exchanges are included. 
Additional subtraction terms
are added to remove the kinematic singularities of multipole amplitudes. 
In solving the dispersion relation equations, only
the $P_{33}$ phases are included to account for the unitarity
condition via Watson Theorem and 
the imaginary parts of other multipole amplitudes are neglected.
Clearly, Adler's approach
needs improvements for a precise extraction of the
axial N-$\Delta$ from factors from the data.
We however will not address this issue in this paper.
Instead, we focus on the development of 
a dynamical approach outlined above.

In section II, we present a formulation for calculating the cross
sections of neutrino induced pion production reactions.
The  current matrix elements needed for
our calculations are given in section III.
The results are presented in section IV. Conclusions and discussions
 on possible future developments will be given in section V.

\section{Cross Section Formula}

The formula for calculating
the cross sections of neutrino induced pion production
reactions have been given in Ref.\cite{adler}.  
In this section, we would like to present
a different formulation which is more closely related to the commonly used
formulation of pion electroproduction, and is more convenient for our
calculations based on a dynamical formulation of the problem.  

We focus on the $\nu + N \rightarrow l +\pi  + N$ reaction, 
where $l$ stands for electron($e$) or muon($\mu)$
and $\nu$ for electron neutrino($\nu_e$) or muon neutrino($\nu_\mu$).
We start with the interaction Lagrangian 
\begin{eqnarray}
L_{int}(x)= \frac{G_F\cos \theta_c}{\sqrt{2}}[l^\mu(x)J_\mu(x) + \mbox{h.c.}],
\end{eqnarray}
where $G_F=1.16637 \times 10^{-5}$ (GeV)$^{-2}$, $cos\theta_c=0.974$,
\begin{eqnarray}
l^\mu(x)&=&\bar{\psi}_l(x)\gamma^\mu(1-\gamma_5)\psi_\nu(x) 
\end{eqnarray}
is the lepton current and
\begin{eqnarray}
J^\mu(x)=V^\mu(x) - A^\mu(x)
\end{eqnarray}
is the hadron current.
The vector current($V^\mu$) and axial vector current($A^\mu$) will be  
constructed later in terms of hadronic degrees of freedom.

The coordinate system for calculating the differential cross 
sections of the
$\nu(p_\nu) + N(p) \rightarrow l(p_l) + \pi(k)+ N(p^\prime)$ reaction
is chosen such that the leptons are on the
x-z plane. The momentum transfer 
$\bm{q} = \bm{p}_\nu - \bm{p}_l$, which is also the
total momentum of the final $\pi N$ system in the laboratory system, is
along the z-axis. The
final $\pi N$ system forms a plane which has
an angle $\phi_\pi$ with respect to the lepton $\nu$-$l$ plane. This choice
is identical to that widely used in the study of pion electroproduction. 
It is then straightforward to show that 
the differential cross section  can be written as
\begin{eqnarray}
\frac{d\sigma^5}{d E_l d\Omega_l d\Omega_\pi^*}
=\frac{G_F^2\cos^2\theta_c}{2}\frac{\ples}{\pnus}
 \frac{\ppisc m_N}{(2\pi)^5 W} L^{\mu\nu}W_{\mu\nu}, \label{cross}
\end{eqnarray}
where $m_N$ is the nucleon mass,
 and the phase space factor is expressed in terms of
lepton momenta $\bm{p}_l$ and $\bm{p}_\nu$ in the laboratory(Lab) frame,
 pion momentum $\bm{k}_c$ in the $\pi N$ center of mass(C.M.) frame, and
the invariant mass $W$ of the $\pi N$ subsystem.
The angle $\Omega_\pi^*$ in Eq.~(\ref{cross}) is the pion scattering 
angle defined in the $\pi N$ center of mass frame. 

The lepton tensor in Eq.~(\ref{cross}) is
\begin{eqnarray}
L^{\mu\nu} = p^\mu_l p^\nu_\nu +p^{\nu}_l p^{\mu}_\nu
  -g^{\mu\nu}(p_\nu\cdot p_l)
 \pm i \epsilon^{\alpha\beta\mu\nu}p_{\nu,\alpha} p_{l,\beta}. \label{lmunu}
\end{eqnarray}
Here $\epsilon^{0123}= 1$ and $+(-)$ in the last term is for neutrino
(anti-neutrino) reactions.
The hadron tensor is
\begin{eqnarray}
W_{\mu\nu} = \frac{1}{2}\sum _{s^\prime_N s_N} j_\mu j^*_\nu,
\end{eqnarray}
where $j_\mu$ implicitly represents the matrix element 
\begin{eqnarray}
< k , p^\prime s_N^\prime \mid J_\mu \mid  p s_N>
 & = & \frac{1}{(2\pi)^3}\sqrt{\frac{m_N^2}{2E_\pi(k) E_N(p')E_N(p)}}j_\mu
\end{eqnarray}
with $s_N$ denoting the
z-component of nucleon spin. Such a simplified notation
for hadron current matrix element will be used in the rest of this
paper.

To see the current non-conserving part of hadron current
 explicitly, we introduce the following variables
\begin{eqnarray}
K^\mu & = & \frac{p^\mu_\nu + p^\mu_l}{2},\\
q^\mu & = & p^\mu_\nu - p^\mu_l, \\
Q^2   & = & - q^{2}  = - m_l^2 + 2 (p_\nu\cdot p_l) \label{q2},
\end{eqnarray}
where $q^\mu$ is the lepton momentum transfer and $m_l$ is the mass of
lepton $l$.
We then obtain
\begin{eqnarray}
L^{\mu\nu}W_{\mu\nu} & = & \frac{1}{2}\sum_{s_N',s_N}[
 2(K\cdot j)(K\cdot j^*) - \frac{(q\cdot j)(q\cdot j^*)}{2}
 - \frac{(j\cdot j^*)(Q^2 + m_l^2)}{2} \nonumber \\
& & \pm i 
\epsilon^{\alpha\beta\mu\nu}q_\alpha K_\beta j_\mu j_\nu^*] \label{lj1}.
\end{eqnarray}

The hadron current $j^\mu$ in Eq.~(\ref{lj1}) is defined in the 
Lab frame.
To account for the final $\pi N$ interaction within the dynamical formulation
and to get an expression in terms of the scattering angle $\Omega^*_\pi$ 
of Eq.~(\ref{cross}),
we need to  write $j^\mu$ in terms of the current $j^\mu_c$ defined
in the $\pi N$ C.M. frame.
With the choice of the coordinate system described above, the transformation
from the  Lab frame to the $\pi N$ C.M. frame 
is just a Lorentz boost along the z-axis. For the
 x and y components of the hadron current, we obviously have
the following simple relations 
\begin{eqnarray}
j^x & = & \cphi j_c^x - \sphi j_c^y \label{jcrot},\\
j^y & = & \sphi j_c^x + \cphi j_c^y.
\end{eqnarray}
For the z and time components, the transformation relations are
\begin{eqnarray}
j^0 & = & \ch   j_c^0 + \sh  j_c^z, \\
j^z & = & \sh  j_c^0 + \ch   j_c^z,
\end{eqnarray}
where
\begin{eqnarray}
\ch & = \frac{E_{N,c}}{m_N} & = \frac{m_N + \omega}{W},\\
\sh  & =  \frac{\pqsc}{m_N} & =  \frac{\pqs}{W}.
\end{eqnarray}
Here we have introduced $q^\mu =( \omega, \bm{q})$ and
$q^\mu_c = (\omega_c, \bm{q}_c)$ as the momentum transfers in the
Lab and $\pi N$ C.M. frames respectively, and
$E_{N,c}=\sqrt{\bm{q}^2_c + m_N^2}$.

Since the leptons are on the x-z plane and the momentum transfer $\bm{q}$
is in the z-axis, we obviously have
\begin{eqnarray}
K^x &=&  \frac{\pnus\ples}{\pqs}\sin\theta_l,\\
K^y &=& 0,\\
K^z &=& \frac{\pnus^2 - \ples^2}{2\pqs},\\
K^0 & = & \frac{\pnus + E_l}{2}. \label{k0}
\end{eqnarray}
In the above equations, $\theta_l$ is the angle between the outgoing lepton
momentum $\bm{p}_l$
and the incident neutrino momentum $\bm{p}_\nu$.

By using the above relations Eqs.~(\ref{jcrot})-(\ref{k0}), it is
 straightforward, although tedious, to cast Eq.~(\ref{lj1}) into the following
form
\begin{eqnarray}
L^{\mu\nu}W_{\mu\nu} & = & \frac{1}{2}\sum_{s_N',s_N}[
 R_T+R_L + R_{LT} \cphi + R_{TT} \cphit \nonumber \\
& & + R_{LT'} \sphi + R_{TT'} \sphit ] \label{lj2}
\end{eqnarray}
All terms in the above equation
 can  be calculated from $j^\mu_c$ defined in the
$\pi N$ C.M. frame. Explicitly, $R_T$ and $R_L$ terms are given as
\begin{eqnarray}
R_T & = & ((K^{x})^2 + \frac{Q^2 + m_l^2}{2})(|j^x_c|^2 + |j^y_c|^2)
               \mp x \mbox{Im}(j^x_c j^{y *}_c), \label{t1t}\\
R_L & = &
 \frac{1}{2\bm{q}_c^2}[(x^2 - Q^4 - Q^2m_l^2)|j^0_c|^2
    + 2(xy - Q^2\omega_c - \omega_c m_l^2)Re(j_c^0 \rho^*) \nonumber \\
  &&  + (y^2 - \omega_c^2 + m_l^2)|\rho|^2], \label{t1l}
\end{eqnarray}
with
\begin{eqnarray}
x & = & \frac{(E_\nu + E_l)Q^2 - \omega m_l^2}{|\bm{q}|}, \\
y & = & \frac{(E_\nu + E_l)\omega_c + \ch \ m_l^2}{|\bm{q}|}. \label{yyy}
\end{eqnarray}
In deriving Eq.~(\ref{t1l}), we have 
defined $\rho=q\cdot j=\omega_c j^0_c - |\bm{q}_c| j^z_c$ 
and eliminated $j^z_c$. 

The other terms in Eq.~(28) are given as
\begin{eqnarray}
 R_{LT}  & = & 
 \frac{2K^x}{\pqsc}( -\mbox{Re}((x j^0_c + y \rho)j^{x*}_c)
                    \pm  \mbox{Im}( (Q^2 j^0_c + \omega_c \rho)j^{y*}_c),\\
R_{TT}  & = &
   (K^{x})^2(|j^x_c|^2 - |j^y_c|^2), \\
R_{LT'} & = & 
 \frac{2K^x}{\pqsc}( \mbox{Re}((x j^0_c + y\rho)j^{y*}_c)
                    \pm  \mbox{Im}
( (Q^2 j^0_c + \omega_c \rho)j^{x*}_c),\\
R_{TT'} & =&  -2(K^{x})^2\mbox{Re}(j^x_c j^{y *}_c).\label{t5}
\end{eqnarray}

It is interesting to note that in the $m_l=0$ limit
the above formula become 
\begin{eqnarray}
R_T & = & \frac{Q^2}{1-\epsilon} [
  \frac{|j^x_c|^2 + |j^y_c|^2}{2}
  \mp \sqrt{1-\epsilon^2}\mbox{Im}(j^x_c j^{y *}_c)], \\
R_L & = & \frac{ Q^2}{1-\epsilon}\epsilon\frac{Q^2}{\pqsc^2}|\bar{j}_c^0|^2,\\
R_{LT} & = &
 \frac{Q^2}{1-\epsilon}
\sqrt{\frac{2\epsilon(1+\epsilon)Q^2}{\pqsc^2}}
{}[  -\mbox{Re}(\bar{j}^0_c j^{x*}_c) 
     \pm \sqrt{\frac{1-\epsilon}{1+\epsilon}}\mbox{Im}(\bar{j}^0_c j^{y*}_c)],
    \\
R_{TT}  & = & \frac{Q^2}{1-\epsilon}[
     \frac{\epsilon(|j^x_c|^2 - |j^y_c|^2)}{2}],\\
R_{LT'} & =&
 \frac{Q^2}{1-\epsilon}
\sqrt{\frac{2\epsilon(1+\epsilon)Q^2}{\pqsc^2}}
{}[  \mbox{Re}(\bar{j}^0_c j^{y*}_c) 
     \pm \sqrt{\frac{1-\epsilon}{1+\epsilon}}\mbox{Im}(\bar{j}^0_c j^{x*}_c)],
\\
R_{TT'} & = &  \frac{Q^2}{1-\epsilon}  [
 -\epsilon \mbox{Re}(j^x_c j^{y *}_c)],
\end{eqnarray}
where
\begin{eqnarray}
\epsilon = \frac{1}{1 + \frac{2 \mid \bm{q}\mid^2}{Q^2} 
\tan^2\frac{\theta_l}{2}} 
\end{eqnarray} 
with $\bar{j}_c^0 = j_c^0 + \omega_c\rho/Q^2$. Eqs.~(37)-(43) are 
very similar to the familiar forms of pion electroproduction.
In the energy region we are considering, the muon mass cannot be 
neglected. Calculations of the differential cross section
Eq.~(\ref{cross}) for the $\nu_\mu + N \rightarrow \mu + \pi + N$ reactions
must be done by using Eqs.~(\ref{t1t})-(\ref{t5}).

To deal with the old data, we need to calculate the differential
cross section in terms of the variables
$W$(invariant mass of $\pi N$ subsystem), $Q^2=-q^2$(Eq.~(\ref{q2}))
instead of lepton energy ($E_l$) and scattering angle $\theta_l$.
The relation between them are
\begin{eqnarray}
Q^2 & = &2E_\nu E_l - 2 |\bm{p}_l||\bm{p}_\nu|\cos\theta_l - m_l^2, \\
W & = & \sqrt{m_N^2 + 2m_N(E_\nu-E_l) - Q^2}.
\end{eqnarray}
By using the above relations, we can obtain the following
relation
\begin{eqnarray}
\frac{d\sigma}{dW dQ^2d\Omega^*_\pi} & = & \frac{2\pi W}{2m_N \pnus \ples}
        \frac{d\sigma}{dE_l d\Omega_l d\Omega^*_\pi}.
 \label{crosswq}
\end{eqnarray}
To calculate the differential cross sections in the right-hand-side of
the above equation using Eq.~(\ref{cross}) and 
Eqs.~(\ref{t1t})-(\ref{t5}), we also need to know the
variables in $\pi N$ C.M. frame.
These are given by
\begin{eqnarray}
\omega_c & = & \frac{W^2 - Q^2 -m_N^2}{2W},\\
|\bm{q}_c| & = & \frac{m_N}{W}|\bm{q}|,\\
|\bm{k}_c| & = & \sqrt{(\frac{W^2+m_\pi^2-m_N^2}{2W})^2 - m_\pi^2}.
\end{eqnarray}

The total cross section for a given incident neutrino energy $E_\nu$
is then calculated by
\begin{eqnarray}
\sigma(E_\nu) & = & \int_{W_{min}}^{W_{max}} dW 
                    \int_{Q^2_{min}}^{Q^2_{max}} d Q^2 \frac{d\sigma}{dW dQ^2},
\label{crosst}
\end{eqnarray}
where 
\begin{eqnarray}
\frac{d\sigma}{dWdQ^2}& =&
 \int d\Omega^*_\pi \frac{d\sigma}{dWdQ^2d\Omega^*_\pi}\nonumber \\
 & = & \frac{G_F^2\cos^2\theta_c}{2}
  \frac{1}{32\pi^4}\frac{\ppisc}{\pnus^2} \int d\Omega^*_\pi
 \frac{1}{2}\sum_{s_N',s_N}[ R_T + R_L].
\label{crosswq1}
\end{eqnarray}
The integration ranges in Eq.~(\ref{crosst}) are found to be
\begin{eqnarray}
W_{min} &=& m_N + m_\pi, \\
W_{max} &=& W_T - m_l,
\end{eqnarray}
where $m_\pi$, $m_N$, and $m_l$ are the masses of
the pion, nucleon and the outgoing lepton respectively, and
 $W_T=\sqrt{(p_\nu + p_N)^2}=
\sqrt{2m_N E_\nu + m_N^2}$ is the invariant mass of
the initial $\nu$-$N$ system.
For a given allowed $W$, the range of $Q^2$ is found to be
\begin{eqnarray}
Q^2_{min} &=& -m^2_l + 2 E^c_\nu(E^c_l-p^c_l), \\
Q^2_{max} &=& -m_l^2 + 2E^c_\nu(E^c_l+p^c_l),
\end{eqnarray}
where $E^c_l=\sqrt{m^2_l+p^{c2}_l}$, $E^c_\nu$ and $p^c_l$ are
the neutrino energy and outgoing lepton momentum in the
C.M. frame of the whole system(not the C.M.  frame
of the final $\pi N$ subsystem).
Explicitly, we find 
\begin{eqnarray}
E^c_\nu &=& \frac{W^2_T-m_N^2}{2W_T}, \\
p^c_l &=& \sqrt{(\frac{W^2_T+W^2-m^2_l}{2W_T})^2-W^2}. \\
\end{eqnarray}

\section{Current Matrix elements}

To proceed, we need to construct current operators $J_\mu = V_\mu- A_\mu$.
The matrix elements of these currents are the input to solving the dynamical
equations that are of the form of Eq.~(2)-(6) with appropriate
changes of notations; namely $v_{\gamma N} \rightarrow v_{V N}$ or
$v_{AN}$ and $\Gamma_{\Delta \rightarrow \gamma N} \rightarrow
\Gamma_{\Delta \rightarrow VN}$ or $\Gamma_{\Delta\rightarrow AN}$.
In the first part of this section, we present the non-$\Delta$ current
matrix elements. The amplitudes associated with the $\Delta$ excitation
will be given in the second subsection.

\subsection{Non-$\Delta$ Amplitudes} 

We first consider the hadronic currents associated with $\pi$ and $N$
degrees of freedom. Here the chiral symmetry is the guiding principle.
These currents can be derived
 by using the well-developed procedures\cite{park,leut}.  
Within the SL model, we need to also consider currents associated with
$\rho$ and $\omega$ mesons.
Since the electromagnetic current($J^{em}_\mu$) is related to 
the isovector vector current($\vec{V}^\mu$) by
$J^{em}_\mu = J^{iso-scalar}_\mu + V^3_\mu$,
we can use the Conserved Vector Current(CVC) hypothesis to obtain the 
weak vector current from $V^3_\mu$ of SL model by isospin rotation.
We find
\begin{eqnarray}
\vec{V}^\mu\cdot\vec{v}_\mu & = & \bar{N}
   {}   [\gamma^\mu - \frac{\kappa^V}{2m_N}\sigma^{\mu\nu}
\partial_\nu\vec{v}_\mu]\cdot
    \frac{\vec{\tau}}{2}N 
 + \frac{f_{\pi NN}}{m_\pi}\bar{N} \gamma^\mu\gamma_5
 \vec{\tau}N \times \vec{\pi}\cdot\vec{v}_\mu\nonumber \\
& &  + [\vec{\pi} \times \partial^\mu \vec{\pi}]\cdot\vec{v}_\mu
  + \frac{g_{\omega\pi V}}{m_\pi}\epsilon_{\alpha\beta\gamma\delta} 
[\partial^\alpha\vec{v}^\beta]\cdot \vec{\pi}
[\partial^\gamma \omega^\delta], 
\end{eqnarray}
where $\vec{v}_\mu$ is an arbitrary isovector function.
Note that $\rho$ meson does not contribute to the charged vector currents
considered in this work since $\rho$-$\pi$ current is isoscalar.

To construct axial vector currents associated with $\pi$ and N 
degrees of freedom, we are guided by
 the standard effective chiral Lagrangian methods\cite{park,leut} and
follow the procedure of SL model.
We then obtain the following form of axial vector current
\begin{eqnarray}
\vec{A}^{\mu} & = & g_A \bar{N} \gamma^\mu \gamma_5\frac{\vec{\tau}}{2} N
     - f_{\rho\pi A}\vec{\rho}^\mu \times \vec{\pi}
     - F \partial^\mu \vec{\pi}.
\end{eqnarray}
Here $F=93$ MeV is the pion decay constant, and $g_A=1.26$ is the
nucleon axial coupling constant.
Note that $\omega$-$\pi$ current is G-parity violating second class 
current and is not considered here.

With the above current operators and the following Lagrangians from
SL model
\begin{eqnarray}
L_{\pi NN} & = & -\frac{f_{\pi NN}}{m_\pi}\bar{N}  \gamma_\mu \gamma_5
             \vec{\tau}N \cdot \partial^\mu  \vec{\pi}, \\
L_{\rho N\pi} & = & g_{\rho} (\bar{N}
   {}   [\vec{\sla{\rho}}
 - \frac{\kappa^\rho}{2m_N}\sigma_{\mu\nu}\partial^\nu \vec{\rho}^\mu]
    \cdot\frac{\vec{\tau}}{2}N + 
  \vec{\pi}\times \partial_\mu \vec{\pi}\cdot \vec{\rho}^\mu), \\
L_{\omega NN} & = & g_{\omega NN}\bar{N}
   {}   [\sla{\omega} - \frac{\kappa^\omega}{2m_N}\sigma_{\mu\nu}\partial^\nu
          \omega^\mu   ]N, 
\end{eqnarray}
we can evaluate the current matrix elements of the tree-diagrams
illustrated in Figs.~1-2. 

For vector current contributions(Fig.~1), we have 
\begin{eqnarray}
V^\mu(k,j,q,i)  & = & \frac{1}{(2\pi)^3}
 \sqrt{\frac{m_N^2}{2E_\pi(k)E_N(p)E_N(p')}} \bar{u}(\bm{p}')[
V^\mu_{Born}(k,j,q,i)+ V^\mu_{\omega,\pi}(k,j,q,i)]u(\bm{p}),\nonumber \\
\end{eqnarray}
where the Born term(Figs.~1(a)-(d)) is
\begin{eqnarray}
V^\mu_{Born}(k,j,q,i)  & = &
 i [ \Gamma_\pi(k,j) S_F(p'+k) V^\mu_N(q,i) +
  V^\mu_N(q,i)  S_F(p-k)\Gamma_\pi(k,j) ] \nonumber \\
& &+  \frac{f_{\pi NN}}{m_\pi}\epsilon_{ijk}\tau_k[
 - \gamma^\mu\gamma^5 - \frac{(\sla{p}'-\sla{p})\gamma^5}{(p-p')^2 - m_\pi^2}
  (k + p - p')^\mu]
\end{eqnarray}
with
\begin{eqnarray}
\Gamma_\pi(k,j)& = & \frac{f_{\pi NN}}{m_\pi}\sla{k}\gamma_5\tau^j, \\
V_N^\mu(q,i) & = & [\gamma^\mu + i \frac{\kappa^V}{2m_N}\sigma^{\mu\nu}q_\nu]
\frac{\tau^i}{2},
\end{eqnarray}
and 
\begin{eqnarray}
S_F(p)=\frac{1}{\sla{p}-m_N} \nonumber
\end{eqnarray}
is the Dirac propagator.
The $\omega$-$\pi$ term (Fig.~1e) is
\begin{eqnarray}
V^\mu_{\omega,\pi}(k,j,q,i)= -\delta_{ij}
\frac{g_{\omega NN}g_{\omega\pi V}}{m_\pi}[\gamma^\alpha
+\frac{i\kappa_\omega}{2m_N}\sigma^{\alpha \beta}(p'-p)_\beta]
 \frac{\epsilon^\mu_{\ \  \beta'\gamma\alpha}q^{\beta'}(p'-p)^\gamma}
    {(p-p')^2 - m_\omega^2}.
\end{eqnarray}
The $\rho-\pi$ term has a similar form, but it is a isoscalar and
does not contribute the charged currents considered in this investigation.

The non-$\Delta$ axial vector current amplitude has two parts.
The first part is due to the first two terms of Eq.~(60) and is
illustrated in Fig.~2. The second part is due to
the term $-F\partial^\mu \vec{\pi}$.  Obviously, this
interaction will induce a pion-pole term  illustrated in Fig.~3, where 
the shaded box is identical to Fig.~2 except that the  
axial vector field(waved line) is replaced by the pion field(dashed line).
For the non-pion pole part of the axial current contributions(Fig.~2), 
we have
\begin{eqnarray}
A_{NP}^\mu(k,j,q,i)  & = & \frac{1}{(2\pi)^3}
\sqrt{\frac{m_N^2}{2E_\pi(k)E_N(p)E_N(p')}} \bar{u}(\bm{p}')[
A^\mu_{Born}(k,j,q,i)+ A^\mu_{\rho,\pi}(k,j,q,i)]u(\bm{p}), \nonumber  \\
\end{eqnarray}
where the Born term is
\begin{eqnarray}
A^\mu_{Born}(k,j,q,i)  & = &
 i [ \Gamma_\pi(k,j) S_F(p'+k) A^\mu_N(q,i) +
  A^\mu_N(q,i)  S_F(p-k)\Gamma_\pi(k,j) ],
\end{eqnarray}
with
\begin{eqnarray}
A_N^\mu(q,i) & = &g_A \gamma^\mu \gamma^5 \frac{\tau^i}{2}.
\end{eqnarray}
The $\rho$-$\pi$ term in Eq.~(69) is
\begin{eqnarray}
A_{\rho,\pi}^{\mu}(k,j,q,i) & = & - g_\rho f_{\rho\pi A}\epsilon_{ijk}\frac{\tau^k}{2}
{}[\gamma^\mu + i \frac{\kappa^\rho}{2m_N}\sigma^{\mu\nu}(p'-p)_\nu ]
\frac{1}{m^2_\rho - (p'-p)^2}.
\end{eqnarray}
Taking a phenomenological point of view, we fix the coupling constant
$f_{\rho\pi A}$ in the above equation by using the soft pion limit
\begin{eqnarray}
f_{\rho\pi A} &= & \frac{m_\rho^2}{F g_\rho}.
\end{eqnarray}

The  pion pole term(Fig.~3) can be easily obtained by modifying 
$A^\mu_{NP}$, defined by Eq.~(69), to
 include a pion propagator. By using the PCAC relation,
we find  that this pion 
pole term can be easily included by the following procedure
\begin{eqnarray}
A^\mu(k,j,q,i) & = &
  A_{NP}^\mu(k,j,q,i) - \frac{q^\mu q\cdot A_{NP}(k,j,q,i)}
                             {q^2 - m_\pi^2} .
\end{eqnarray}

In the dynamical approach, as briefly reviewed in section I, the non-resonant
amplitudes will be integrated over the $\pi N$ scattering amplitudes. 
Thus the non-$\Delta$ amplitudes, as illustrated in Figs.~1-2, must be
regularized by introducing a form factor at each vertex. Furthermore, the
finite size effects of hadron structure also require including form factors. 
Fortunately, these form factors
can be taken from the SL model and other theoretical investigations.
The vector current matrix elements are regularized as those in SL model.
For axial current matrix elements, the $A$-$NN$ vertex is
regularized by a dipole form factor
\begin{eqnarray}
F_D(Q^2)=\frac{1}{(1 + Q^2/M_A^2)^2} , \label{axialf}
\end{eqnarray}
where $M_A = 1.02$ GeV is taken from Ref.\cite{meissner}. The $\pi$-$NN$
and $\rho$-$NN$ form factors are taken from SL model\cite{sl1}.
The $A$-$\rho$-$\pi$ form factor is also assumed to be of the dipole form of 
Eq.~(\ref{axialf}).
With these specifications, the non-$\Delta$ amplitudes do not have
any adjustable parameters in our investigations.

\subsection{$\Delta$ amplitudes}

The $\Delta$ has two contributions to weak pion production reactions,
as shown in Fig.~4. The resonant amplitude is due to the formation of
a $\Delta$ in s-channel(Fig.~4a). The cross $\Delta$ term, Fig.~4b, is
part of the non-resonant amplitude. Each term has a corresponding
pion pole term illustrated in Fig.~3 and these pion pole terms must be also
included in our investigations.
  
To calculate
these $\Delta$ amplitudes, we need to define the matrix 
elements $<\Delta \mid V^\mu \mid N >$ and 
$< \Delta \mid A^\mu \mid N >$. The vector current matrix element
$<\Delta \mid V^\mu \mid N >$ can be obtained from SL model by
appropriate isospin rotations. Here we focus on the axial vector current 
matrix  element.

It is well known\cite{adler,hhm} that the most general
form of axial vector current matrix element can be written as
\begin{eqnarray}
<\Delta|A^{\mu i} |N>  = \bar{u}_{\Delta \nu}({\bm p}')
\Gamma^{\mu\nu}_A
T^i u({\bm p})  \label{delta1}
\end{eqnarray}
with
\begin{eqnarray}
\Gamma^{\mu\nu}_A&=&
 d_1(q^2) g^{\mu\nu}+
   \frac{d_2(q^2)}{m_N^2}P_{\alpha}(q^\alpha g^{\mu\nu}-q^\nu g^{\alpha\mu})
  - \frac{d_3(q^2)}{m_N^2}p^\nu q^\mu 
+ i\frac{d_4(q^2)}{m_N^2}\epsilon^{\mu\nu\alpha\beta}
P_\alpha q_\beta \gamma_5 ] \nonumber 
\end{eqnarray}
where $u_{\Delta \nu}({\bm p})$ is the spin $3/2$ Rarita-Schwinger spinor,
$q=p'-p$, $P=p'+p$, and $T^i$ is the $i-$th component of the
isospin transition operator(defined by the reduced matrix element
$<3/2 \mid\mid {\bm T} \mid \mid 1/2> = 
-<1/2 \mid\mid {\bm T}^+ \mid\mid 3/2 > = 2$ in Edmonds convention\cite{edmonds}).

It is useful to explore here the meaning of each form factor of
Eq.~(\ref{delta1}). 
 Since $u_{\Delta\nu}({\bm p}') = u({\bm p}')\epsilon_{\Delta,\nu}
({\bm p}')$, where $\epsilon_{\Delta,\nu}({\bm p}')$
 is an unit  four vector, we can rewrite Eq.~(\ref{delta1}) as
\begin{eqnarray}
<\Delta|A^{\mu i} |N> & = &
\bar{u}({\bm p'})[
 (d_1(q^2) + \frac{P\cdot q}{m_N^2}d_2(q^2))\epsilon_\Delta^\mu
 + (-d_2(q^2) P^\mu + d_3(q^2) q^\mu)\frac{q\cdot\epsilon_\Delta}{m_N^2}
 \nonumber \\& &
+ i\frac{d_4(q^2)}{m_N^2}\epsilon^{\mu\nu\alpha\beta}
\epsilon_{\Delta,\nu}P_\alpha q_\beta \gamma_5]T^i u({\bm p}). 
\end{eqnarray}
In the rest frame of a $\Delta$ on the resonance energy(
 $p' =  (m_\Delta, \bm{0}),  p   =  (E_N(q),-\bm{q}),
 q = (m_\Delta-E_N(q), \bm{q}), 
\epsilon_\Delta  =  (0,\bm{\epsilon})$, $\bar{u}(p'=(m_\Delta,0))
=(\chi^+_\Delta,0)$, and $u(p) \propto \chi_N$, where $\chi_\beta$ is the Pauli
spinor), the space and time components of
Eq.~(77) become 
\begin{eqnarray}
<\Delta|\bm{A}^i |N> & = & 
\sqrt{\frac{E_N+m_N}{2m_N}}[
(d_1 + \frac{m_\Delta^2-m_N^2}{m_N^2}d_2)\bm{S}
 - (d_2 + d_3)\frac{(\bm{S}\cdot\bm{q})\bm{q}}{m_N^2} \nonumber \\
& & -id_4 \frac{\bm{S}\times \bm{q}(\bm{\sigma}\cdot\bm{q})}
         {m_N^2(E_N+m_N)}] T^i ,\\
<\Delta|A^{0 i} |N> & = & 
\sqrt{\frac{E_N+m_N}{2m_N}}[
 d_2 \frac{\bm{S}\cdot\bm{q}(m_\Delta+E_N)}{m_N^2}
-d_3  \frac{\bm{S}\cdot\bm{q}(m_\Delta-E_N)}{m_N^2}]T^i ,
\end{eqnarray}
where we have defined the transition spin 
$\bm{S}=\chi^+_\Delta \bm{\epsilon}\chi_N$ (the reduced
matrix element  $<\frac{3}{2} \mid\mid {\bm S} \mid\mid \frac{1}{2}>$ 
is identical to that of the transition isospin $\bm{T}$).
The above expression suggests that
$d_1,d_2$ terms describe the Gamow-Teller transition 
and $d_4$ describes the quadrupole transition.
The term of $d_3$ comes from the pion pole term illustrated in Fig.~3.

For simplicity, we now follow Ref.\cite{hhm} to fix the form factors $d_i(q^2)$ at $q^2=0$
 using the non-relativistic constituent quark model. The axial vector
current  operator for a
constituent quark is derived from taking the
non-relativistic limit of  the standard form
$g_{Aq} \bar{q}\gamma^\mu \gamma_5\frac{\tau}{2} q$. By using a procedure
similar to Eq.~(74) based on the PCAC relation, we then can extend the
resulting current operator to
also include the pion pole term 
contribution(Fig.~3) induced by
the current $-F\partial^\mu \vec{\pi}$ of Eq.~(60). 
The total axial vector current
operators associated with a constituent quark model are found to be
of the following forms
\begin{eqnarray}
<\bm{p}^\prime \mid \bm{A}^i \mid \bm{p}>
&=& g_{Aq}[\bm{\sigma} + \frac{\bm{q}}{q^2-m_\pi^2}
\bm{\sigma}\cdot\bm{q}]\frac{\tau^i}{2}, \\
<\bm{p}^\prime \mid A^{0 \ i} \mid \bm{p} >
&=&g_{Aq}[\frac{\bm{\sigma}\cdot (\bm{p}+\bm{p}^\prime)}{2m_q}
+\frac{q^0}{q^2-m_\pi^2}\bm{\sigma}\cdot \bm{q}]\frac{\tau^i}{2}, 
\end{eqnarray}
where $\bm{p}$ is the quark momentum, $\sigma$ and $\tau$ are the quark
spin and isospin operators respectively.
By using the standard non-relativistic s-wave quark wave functions for
$N$  and $\Delta$, it is
straightforward to obtain
\begin{eqnarray}
<\bm{p}_N^\prime \mid \bm{A}^i \mid \bm{p}_N > 
&=&(\frac{5}{3}g_{Aq})
[\bm{\sigma} + \frac{\bm{q}}{q^2-m^2_\pi}\bm{\sigma}\cdot\bm{q}] 
\frac{\tau^i}{2} ,
 \\
<\bm{p}_N^\prime \mid A^{0 \ i} \mid \bm{p}_N > 
&=&(\frac{5}{3}g_{Aq})
[\frac{ \bm{\sigma}\cdot(\bm{p}_N^\prime+\bm{p}_N)}{2 (3m_q)}
+\frac{q^0}{q^2-m_\pi^2}\bm{\sigma}\cdot\bm{q}] 
\frac{\tau^i}{2}, 
\end{eqnarray}
and
\begin{eqnarray}
<\bm{p}_\Delta \mid \bm{A}^i \mid \bm{p}_N > 
&=&\frac{1}{2}\sqrt{\frac{72}{25}}(\frac{5}{3}g_{Aq})
[\bm{S} + \frac{\bm{q}}{q^2-m^2_\pi}\bm{S}\cdot\bm{q}] T^i , \\
<\bm{p}_\Delta \mid A^{0 \ i}\mid \bm{p}_N > 
&=&\frac{1}{2}\sqrt{\frac{72}{25}}(\frac{5}{3}g_{Aq})
[\frac{\bm{S}\cdot(\bm{p}_\Delta+\bm{p}_N) }{2 (3m_q)}
+\frac{q^0}{q^2-m_\pi^2}\bm{S}\cdot\bm{q}] T^i .
\end{eqnarray}
In above equations, ${\bm p}_N$ and ${\bm p}_\Delta$ are the momenta for 
$N$ and
$\Delta$ respectively, and we have expressed the right hand sides in terms of
usual nucleon spin and isospin operators and the $\Delta$-$N$
transition operators $\bm{S}$ and $\bm{T}$(as defined above).

If we set $3m_q=m_N$ and
$\frac{5}{3}g_{Aq}=g_A(0)=1.26$, 
Eqs.~(82) and (83)
can be identified with the usual non-relativistic form for the
nucleon axial current matrix element at $q^2=0$.
 This then fixes the value of
$g_{Aq}$ which also determine Eqs.~(84)-(85) for
the axial N-$\Delta$ transitions.

Comparing Eqs.~(84)-(85) evaluated at the $\Delta$ rest 
frame($\bm{p}_\Delta=0$) and Eqs.~(78)-(79) taken in the
non-relativistic limit($E_N\rightarrow m_N$ 
and $q^2/m_N^2 \rightarrow 0$), we find that
\begin{eqnarray}
d_1(Q^2_0)&=&g^*_A(Q^2_0) ( 1 +\frac{m^2_\Delta - m^2_N}{2m_N(m_\Delta + m_N)}) \\
d_2(Q^2_0)&=& - g^*_A(Q^2_0)\frac{m_N}{2(m_\Delta + m_N)} \\
d_3(Q^2_0)&=&-g^*_A(Q^2_0)\frac{m^2_N}{q^2-m_\pi^2}
\end{eqnarray}
where $g_A^*(Q^2_0)=\frac{1}{\sqrt{2}}\frac{6}{5}g_A$.
Here we also follow
Ref.\cite{hhm} to assume that the quark model results Eqs.~(84)-(85)
are for the momentum transfer $q_0=(m_\Delta - m_N, \bm{0})$
and $Q^2_0=-q^2_0=(m_\Delta - m_N)^2$.
Eqs.~(86)-(88) agree with the results of Ref.\cite{hhm} if we neglect the  
difference between $m_N$ and $m_\Delta$.

To account for the $q^2$-dependence, we assume that
\begin{eqnarray}
d_i(Q^2)=d_i(0)F(Q^2)
\end{eqnarray}
where $F(Q^2)$ will be specified in the next section. For a given
choice of $F(Q^2)$, we can use Eqs.~(86)-(88) to obtain
$d_i(0)=d(Q^2_0)/F(Q^2_0)$. The
form factors $d_i(Q^2)$ for $i=1,2,3$ 
of the covariant form Eq.(76), which is
used in our numerical calculations, are then completely fixed by
the constituent quark model calculation.
We neglect the $\Delta$ deformation in this work and
set $d_4(q^2)=0$.

With the axial N-$\Delta$ vertex specified above and the $\pi N$ model
constructed in Ref.\cite{sl1}, the dressed $AN\rightarrow \Delta$ vertices
(i.e. Eqs.~(4)-(5)) can be calculated for evaluating the resonant
amplitude Fig.~4a(i.e. the second term of Eq.~(2)).
With the $\pi N\Delta$ 
interaction Lagrangian 
$L_{\pi N\Delta}=f_{\pi N\Delta}/m_\pi \bar{\psi}^\mu_\Delta(x)
\vec{T}N(x)\cdot \partial_\mu \vec{\pi}(x)$ of SL model, 
the cross $\Delta$ term (Fig.~4b) is found to be
\begin{eqnarray}
A^\mu_{\Delta E}(k,j,q,i)=\frac{1}{(2\pi)^3 }
\sqrt{\frac{m_N^2}{2E_\pi(k)E_N(p)E_N(p')}}\bar{u}({\bm p}') 
I^\mu_{\Delta E}(k,j,q,i) u({\bm p}),
\end{eqnarray}
with
\begin{eqnarray}
I^\mu_{\Delta E}(k,j,q,i)=i\frac{f_{\pi N\Delta}}{m_\pi}T^+_i
\Gamma_{A}^{\dagger \mu\nu}S_{\nu\delta}(p-k)T_j k^\delta .
\end{eqnarray}
where the Rarita-Schwinger propagator $S_{\mu\nu}(p)$ 
given explicitly in Eq.~(3.18) of Ref.\cite{sl1}.

The above derivations allow us to use the constituent quark model to
calculate the $\Delta$ amplitudes. This completes our derivations of
the weak current matrix elements which are the input to the dynamical
equations Eqs.~(2)-(4) with the sub-index $\gamma$ replaced by either
 $V$ or $A$.

\section{Results and Discussions}

As presented in subsection III.A, 
the non-$\Delta$ amplitudes, illustrated
in Figs.~1-2, do not have any adjustable parameters in this investigation.
While more theoretical investigations may be needed to
refine these non-resonant amplitudes, it may not be necessary at this
time since the weak pion production data are still limited.
Instead, we focus on the investigation of the axial vector
N-$\Delta$ transition form factors defined in subsection III.B.

To proceed, we first recall the N-$\Delta$ form factors introduced in
SL model. For the bare  magnetic M1 N-$\Delta$ transition, it was taken
as
\begin{eqnarray}
G_M(Q^2)=G_M(0)R_{SL}(Q^2)G_D(Q^2) ,
\end{eqnarray}
where $G_D(Q^2)=1/(1+Q^2/M^2_V)^2$ with $M^2_V= 0.71$ GeV$^2$ is the
usual dipole form factor of the nucleon, and the correction factor
is defined as
\begin{eqnarray}
R_{SL}=(1+aQ^2)\exp(-bQ^2).
\end{eqnarray}
By fitting the pion photoproduction and
electroproduction data up to $Q^2=4$ (GeV/c)$^2$,
it was found that $G_M(0)=1.85$, 
$a=0.154$ (GeV/c)$^{-2}$ and $b=0.166$ (GeV/c)$^{-2}$. 
Here we remark that if we 
follow the same procedure given in section III.B to also calculate the
$\gamma N\rightarrow \Delta$ transition using non-relativistic quark model, we find 
that the quark model yields $G^{Q.M.}_M(Q^2_0)=2.37$ for
$Q^2_0=(m_\Delta-m_N)^2$. This value is very close to the value
$G_M(Q^2_0)=2.42$ which can be  obtained from SL model by using Eqs.~(92)-(93).
This suggests that the use of Eqs.~(86)-(88) for defining
$d_i(Q^2_0)$ of the axial N-$\Delta$ form factors
is a  good starting point of
our investigations. 
As a continuation of SL model, we therefore first consider
a model, called Model I, which assumes that the form factor of Eq.~(89) is
\begin{eqnarray}
F(Q^2)=R_{SL}(Q^2)G_A(Q^2),
\end{eqnarray}
where $R_{SL}(Q^2)$ is given in Eq.~(93), and
$G_A(Q^2)=1/(1+Q^2/M^2_A)^2$
with $M_A=1.02$GeV of the nucleon axial form factor\cite{meissner}.

We first compare the total cross sections predicted by Model I with the
data\cite{data79}. The results are shown in Fig.~5.
We see that the predictions(solid curves)  agree reasonably
well with the data for three pion channels.
 For the data on neutron target, our predictions(solid curves in 
the middle and lower figures)
are in general lower than the data. This is perhaps related to the 
procedures used in Ref.\cite{data79}
to extract these data from the experiments on
deuteron target. 

One of the main features of the dynamical approach taken in this work is
to renormalize(dress) the N-$\Delta$ transitions with the dynamical
pion cloud effect, as described Eqs.~(4)-(5). The importance of
this effect is shown in Fig.~5.
We see that our full calculations(solid curves) are
reduced significantly to dotted curves
if we turn off the dynamical pion cloud effects.
If we further turn off the contributions 
from bare N-$\Delta$ transitions, 
we obtain the dashed curves which correspond to
the contributions from the non-resonant amplitudes(Figs.~1-3).
Clearly, the nonresonant amplitudes are weaker, but
are also essential in getting the good agreement with the data since
they can interfere with the resonant amplitudes.

We next compare the $Q^2$-dependence of the
differential cross sections 
$d\sigma/dQ^2$ predicted by Model I with the data from ANL\cite{data79}
, BNL\cite{data90}, and CERN\cite{data89}. 
Here we need to account for
the variation of neutrino flux in the experiments at ANL\cite{data79}
and BNL\cite{data90}. We calculate the following quantity
\begin{eqnarray}
\frac{d\bar{\sigma}}{dQ^2}=[\int_{E_{min}}^{E_{max}} d E_\nu 
\frac{N(E_\nu)}{\sigma_{model}(E_\nu)}\frac{d\sigma_{model}}{dQ^2}(E_\nu)]/[
 \int_{E_{min}}^{E_{max}} d E_\nu \frac{N(E_\nu)}{\sigma_{model}(E_\nu)}]
\end{eqnarray}
where $N(E_\nu)$ is the distribution of events in neutrino
energy $E_\nu$ which is within the range
 between $E_{min}$ and $E_{max}$, and $\sigma_{model}(E_\nu)$ is the
calculated total cross section.
The distributions $N(E_\nu)$ are given in Fig.~6 of Ref.\cite{data79}
and Fig.~4 of Ref.\cite{data90}.

The predictions from Model I are compared with the 
ANL data\cite{data79} in Fig.~6.
We see that our predictions(solid curve) agree reasonable
well with the data both in magnitude and $Q^2-$dependence.
In Fig.~6 we also compare the contributions
from vector current(dot-dashed curve) and axial vector 
current(dotted curve). They have rather different $Q^2$-dependence
in the low $Q^2$ region and interfere constructively with
each other to yield the solid curve of the full results. 
Since vector current contributions are very much constrained by
the $(e,e'\pi)$ data, the results of Fig.~6 suggest that the
axial vector currents constructed in section III 
are consistent with the data.

For comparing with the BNL data\cite{data90},
we normalize the calculated $d\bar{\sigma}/dQ^2$  
to the events data of Fig.~5 of Ref.\cite{data90} 
at $Q^2=0.2$ (GeV/c)$^2$. As shown in Fig.~7, the data
can be reproduced very well by Model I.
The BNL data was used in the most
 recent  attempt\cite{data90} to extract the axial 
N-$\Delta$ form factor. We will discuss this later.

The comparison with CERN data\cite{data89} is given in Fig.~8.
Here the data have some structure, which is perhaps mainly due to
the poor statistics of experiment. We see that
model I can reproduce the main feature of the $Q^2$-dependence.

To explore the effects due to the dynamical pion cloud on the
 N-$\Delta$ form factors, it is instructive to recall here 
the results of Ref.\cite{sl2} for the magnetic 
M1 $\gamma N\rightarrow \Delta$ transition.  This result 
 is displayed in the left panel of Fig.~9.  
We see that the pion cloud effect
is essential in explaining the empirical values extracted directly
from the data.  The corresponding
dynamical pion cloud effect on the axial N-$\Delta$ form factor is
shown in the right panel of Fig.~9. 
We again see fairly sizable
contribution from dynamical pion cloud. We stress that the empirical
form factor, such as that extracted\cite{data90} from BNL data,
can only be compared
with our dressed form factor, since the dressed form factor, not
the bare form factor, directly determines the reaction amplitude
Eq.~(2). The differences between the solid and dotted curves in Fig.~9 
explain the observation\cite{hhm}
that the quark model prediction of axial vector N-$\Delta$ strength
at $Q^2\rightarrow 0$ is
lower than the empirical value of Ref.\cite{data90}  by about 35 $\%$.
This is simply due to the fact that the dynamical pion cloud is not included
in constituent quark model calculation and most of the hadron model
calculations of N-$\Delta$ transitions. 
It is also interesting to observe from Fig.~9 that the pion cloud effect on the
axial N-$\Delta$ form factor is mainly to increase the magnitude,
not much to change the slope. On the other hand, both the
magnitude and slope of the magnetic M1 form factor
(left-hand side of Fig.~9) are significantly changed by including
the pion cloud effects.

To see the dependence of our predictions on the parameterization
of axial N-$\Delta$ form factor,
we next consider Model II which differs from Model I only in
replacing the correction factor $R_{SL}$ of Eq.~(94) by a different 
form. Rather arbitrarily, we consider a form used in
Refs.\cite{adler,data90,hhm}. The axial N-$\Delta$
form factor in this model II is also defined by Eq.~(89) with
\begin{eqnarray}
F(Q^2)=R_{II}(Q^2)G_A(Q^2)
\end{eqnarray}
where $G_A(Q^2)$ is the axial nucleon form factor of Eq.~(94) and
\begin{eqnarray}
R_{II}=(1+a\frac{Q^2}{b+Q^2})
\end{eqnarray}
with $a=-1.21$ and $b=2$ (GeV/c)$^2$.

The main difference between Model I and Model II is in determining
$d_i(0)=d(Q^2_0)/F(Q^2_0)$ from the quark model values $d_i(Q^2_0)$
given in Eqs.~(86)-(88). We find that $1/F(Q^2_0)=1.20$ for Model I and
$1/F(Q^2_0)=1.26$ for Model II. Thus these two models are already
different at $Q^2\rightarrow 0$. 
Their $Q^2$-dependence are compared in Fig.~10.
We see that the form factor of Model II(dotted curve) are 
much lower than
that of Model I(solid curve). 
 Consequently, the  weak pion production
cross sections calculated from Model II are found to be about 20 $\%$
lower than the data. This is illustrated in Fig.~11.
We find that the model II can fit the data better if we 
increase the strengths
$d_i(0)$ of axial N-$\Delta$ form factor by 20$\%$. The results shown in
Figs.~10 and 11 demonstrate the sensitive of the data to
the form of the bare axial N-$\Delta$ form factor. The data clearly favor
the form factor Eq.~(94). 

As seen in Eqs.~(2) and (4), the empirical N-$\Delta$ form factor extracted
directly from the data, such as those obtained in Ref.\cite{data90} can only
be compared with our dressed form factor $\bar{\Gamma}$. The comparison is 
given in Fig.~12. We see that they do not agree well, in particular
in the high $Q^2$ region. The differences mainly come from the
fact that the non-resonant amplitudes of the Adler Model, which
was used\cite{data90} in extracting the empirical  N-$\Delta$ form factors, 
are rather different from what we have in our dynamical model. With the very
limited data, it is not possible to exclude one of these two
rather different results. 
Clearly, more
precise data are needed for a better test of our models.

To stimulate future experimental efforts, we now present two
predictions of Model I in Figs.~13 and 14.
In Fig.~13, we present the  $Q^2$-dependence of differential cross sections
at $W=1.1, 1.2, 1.236, 1.3$ GeV.
The $W-$dependence
at $Q^2=0.1, 0.5, 1.0, 1.5$ (GeV/c)$^2$ is shown in Fig.~14.  
\section{Conclusions}

We have extended the dynamical model of Refs.\cite{sl1,sl2} to investigate
weak pion production reactions in the $\Delta$ region. 
The calculations for neutrino-induced reactions have been
performed by using the current matrix elements illustrated in Figs.~1-3.
The parameters for the non-resonant terms(Figs.~1-2) are taken from the 
previous
investigations of pion electroproduction. The bare N-$\Delta$
transitions at $Q^2\rightarrow 0$ are fixed by the constituent quark model.
With the axial N-$\Delta$ form factor given in Eqs.~(86)-(88), 
the predicted total cross sections and the differential
cross sections $d\sigma/dQ^2$ agree reasonably well with the existing data. 

We have analyzed the calculated N-$\Delta$ form factors. Similar to
the finding of the $(e,e'\pi)$ studies\cite{sl1,sl2}, we show that the
constituent quark model prediction of axial vector N-$\Delta$ form factor
at $Q^2\rightarrow 0$ is consistent with the data. The discrepancy observed in
Ref.\cite{hhm} is due to the dynamical pion cloud effect which is not
included in the constituent quark model calculation.

An unsatisfactory aspect of the present investigation is the lack of
extensive and good quality data. 
Thus the axial vector  N-$\Delta$ from factor extracted\cite{data90}
from the data relied heavily on model assumptions and hence the 
origins of the differences seen in Fig.~12 are not clear.
It will be very useful to have sufficient
data for performing partial wave decomposition,
 like what have been routinely performed in $(e,e'\pi)$ studies,
such that the N-$\Delta$ form factor can be extracted model independently.
Hopefully, the situation will be improved in the near future when new
neutrino facilities will become available.

\acknowledgments
This work was supported by U.S. DOE Nuclear Physics Division,
Contract No. W-31-109-ENG and Japan Society for the Promotion
of Science, Grant-in-Aid for Scientific Research (C) 12640273.

\vfill
\newpage

\begin{figure}[ht!]
\centering
\includegraphics[width=5.3in]{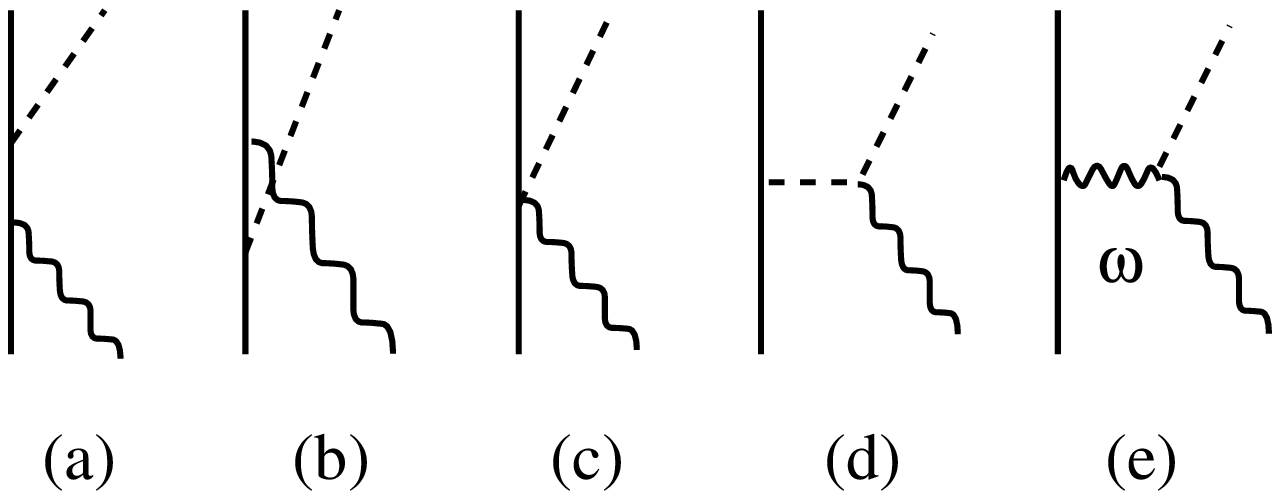}
\caption{Non-$\Delta$ vector current contributions. Dashed lines represent
pion. Waved lines represent the vector field.}
\label{fig:fig1}
\end{figure}

\begin{figure}[ht!]
\centering
\includegraphics[width=3.3in]{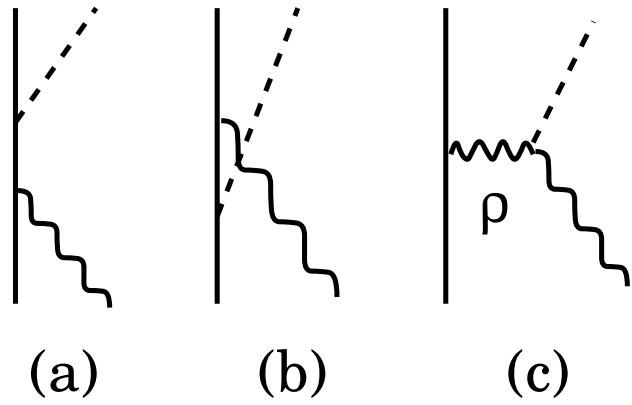}
\caption{Non-$\Delta$ axial vector current contributions. Dashed lines represent
pion. Waved lines represent axial vector field.}
\label{fig:fig2}
\end{figure}

\begin{figure}[ht!]
\centering
\includegraphics[width=1.0in]{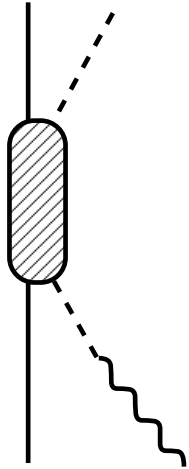}
\caption{Pion-pole term of axial vector current contributions.
The dashed-box represents the either the non-$\Delta$ mechanisms
of Fig.2 or the crossed $\Delta$ mechanism of Fig.4(b).}
\label{fig:fig3}
\end{figure}

\begin{figure}[ht!]
\centering
\includegraphics[width=2.6in]{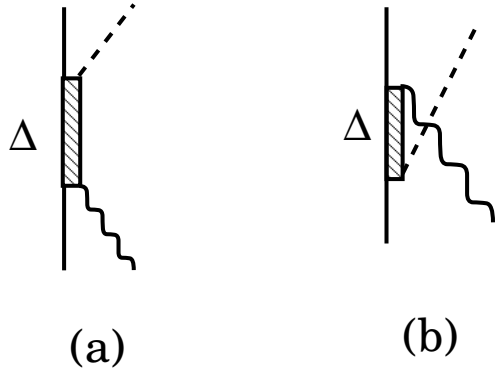}
\caption{The $\Delta$  current contributions. Dashed lines represent pion.
Waved lines represent either the vector field or axial vector field.}
\label{fig:fig4}
\end{figure}

\newpage
\begin{figure}[ht!]
\centering
\includegraphics[width=8cm]{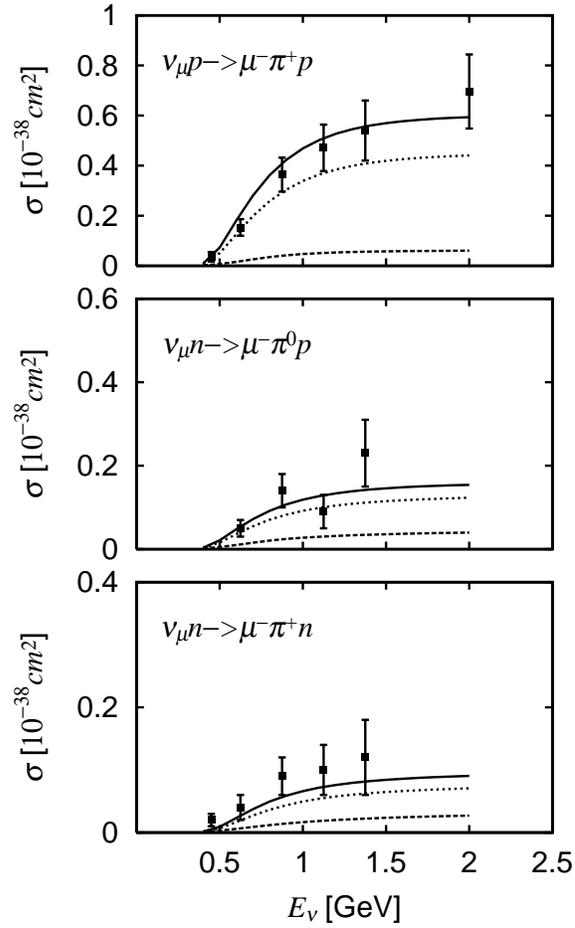} 
\caption{Total cross sections of $N(\nu_\mu,\mu^-\pi)N$
reactions predicted by Model I. 
The data are from Ref.\cite{data79}. The solid curves are from
full calculations. The dotted curves are from turing off pion cloud
effects on N-$\Delta$ transitions. 
The dashed curves are the contributions 
from the non-resonant amplitude.  }
\label{fig:fig9}
\end{figure}

\newpage
\begin{figure}[ht!]
\centering
\includegraphics[width=5.0in]{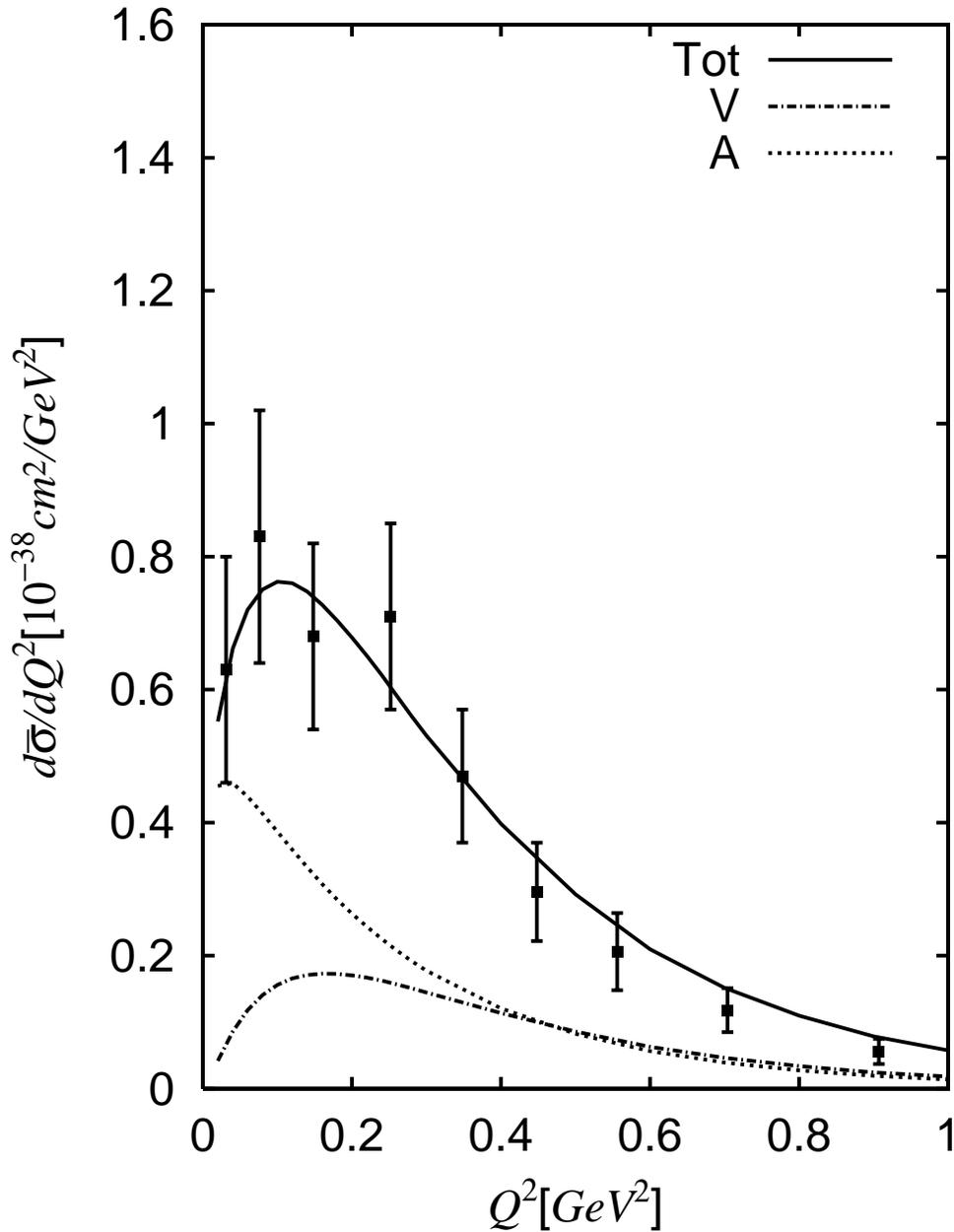}
\caption{Differential cross sections $d\bar{\sigma}/dQ^2$
of  $p(\nu_\mu,\mu^-\pi^+)p$ reaction averaged over 
neutrino energies 0.5 GeV $ < E_\nu < $ 6 GeV.
The curves are from calculations using Eq.~(95).
The dotted 
curve(dot-dashed curve) is the contribution from axial vector 
current A (vector curent V). The solid curve is from our full
calculations with V-A current.
The data are from Ref.\cite{data79}.}
\label{fig:fig5}
\end{figure}

\newpage
\begin{figure}[ht!]
\centering
\includegraphics[width=5.0in]{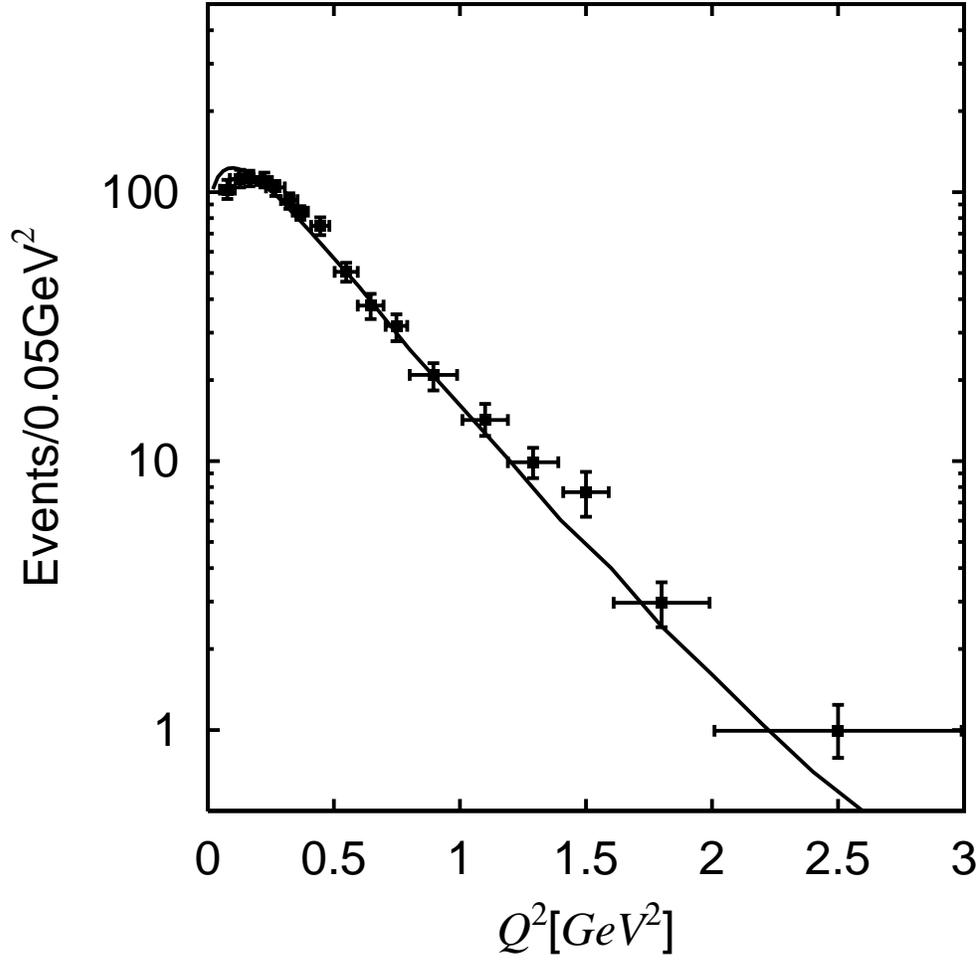} 
\caption{The differential cross sections $d\bar{\sigma}/dQ^2$ of
$p(\nu_\mu,\mu^-,\pi^+)$ reaction calculated
using Eq.~(95). The data are from Ref.\cite{data90}.}
\label{fig:fig7}
\end{figure}

\newpage
\begin{figure}[ht!]
\centering
\includegraphics[width=5.0in]{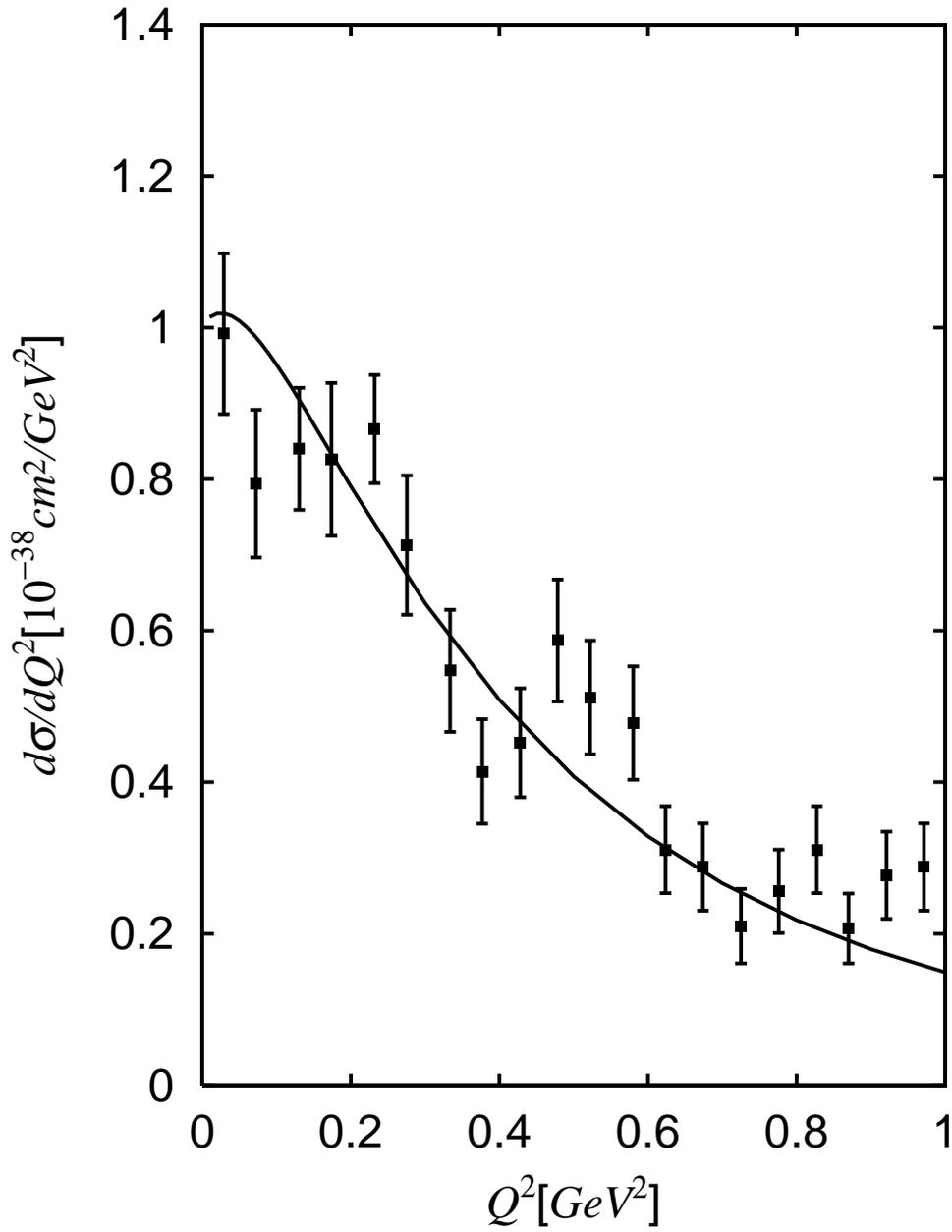} 
\caption{Differential cross sections $d\sigma/dQ^2$
of  $p(\nu_\mu,\mu^-\pi^+)p$ reaction  at
neutrino energy  $ E_\nu= $ 15 GeV. 
 The data are from Ref.\cite{data89}}
\label{fig:fig6}
\end{figure}

\newpage
\begin{figure}[ht!]
\centering
\includegraphics[width=8cm]{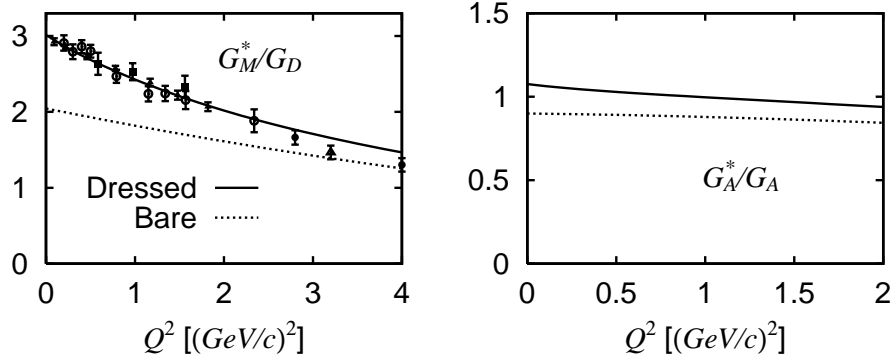} 
\caption{The N-$\Delta$ form factors predicted by Model I:
left panel: Magnetic M1 form factors given in Ref.\cite{sl2}, right panel:
axial vector form factor. The solid curves are from full calculations.
The dotted curves are obtained from turing off the pion cloud
effects. $G_D=1/(1+Q^2/M^2_V)^2$  with $M_V=0.84$ GeV 
is the usual proton form factor and $G_A=1/(1+Q^2/M^2_A)^2$ with $M_A=1.02$ 
GeV is the axial nucleon form factor of Ref. \cite{meissner}.}
\label{fig:figr8}
\end{figure}

\newpage
\begin{figure}[ht!]
\centering
\includegraphics[width=8cm]{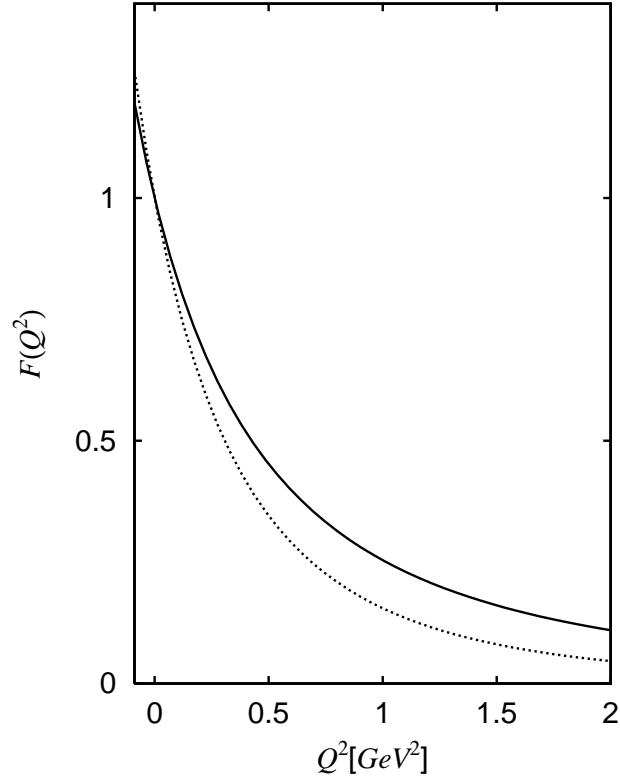} 
\caption{ Bare axial N-$\Delta$ form factors for Model I(soild curve)
and Model II(dotted curve).}
\label{fig:fig10}
\end{figure}

\begin{figure}[ht!]
\centering
\includegraphics[width=8cm]{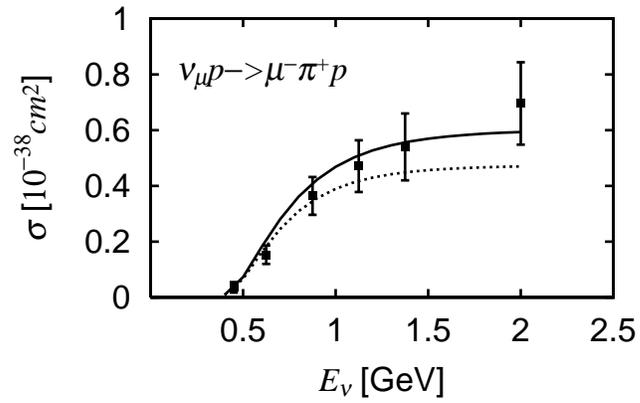} 
\caption{The total cross sections of $p(\nu_\mu,\mu^-\pi^+)p$ calculated
from Model I(soild curves) and Model II(dot curves).}
\label{fig:fig11}
\end{figure}

\begin{figure}[ht!]
\centering
\includegraphics[width=8cm]{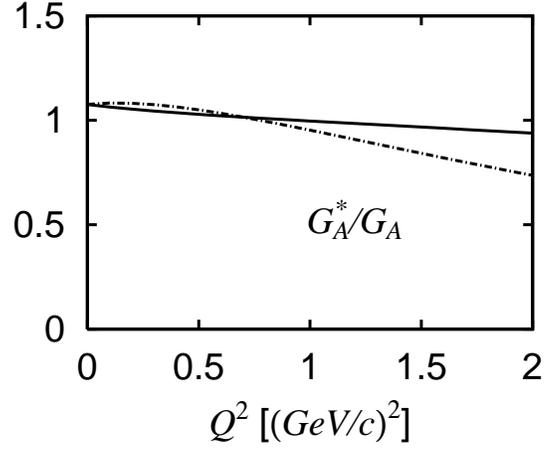} 
\caption{Compare the dressed axial N-$\Delta$ form factor predicted by
Model I(solid curve) with the empirical form factor(dot-dash curve) 
determined in Ref.\cite{data90}.}
\label{fig:fig13}
\end{figure}

\begin{figure}[ht!]
\centering
\includegraphics[width=8cm]{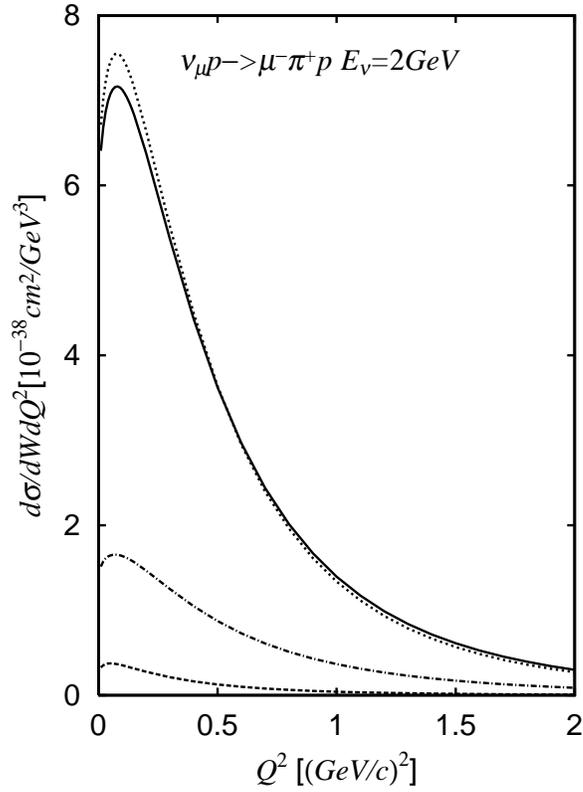} 
\caption{The predicted differential cross sections
$d\sigma/dWdQ^2$ for $p(\nu_\mu,\mu^-\pi^+)p$ at neutrino energy
$E_\nu = 2$ GeV and invariant mass W=1.1(dashed), 1.2(dotted),
1.236(solid), 1.3(dot-dashed) GeV.}
\label{fig:fig14}
\end{figure}

\begin{figure}[ht!]
\centering
\includegraphics[width=8cm]{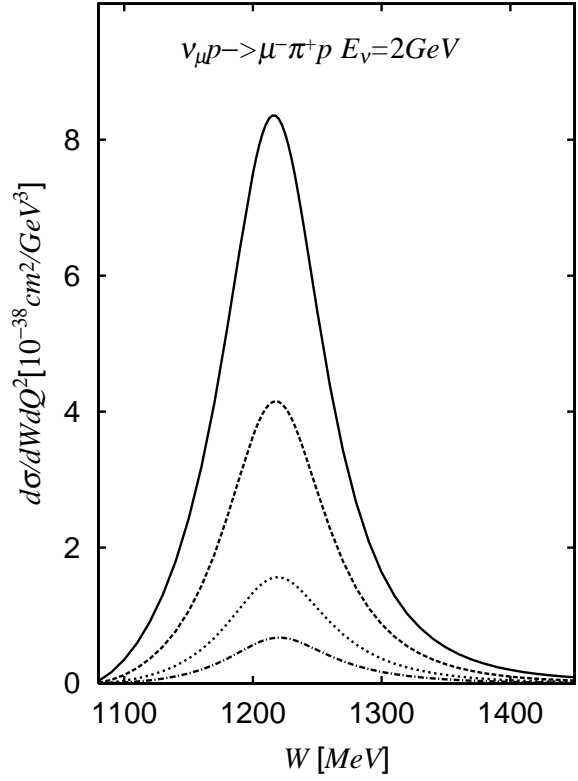} 
\caption{The predicted differential cross sections
$d\sigma/dWdQ^2$ for $p(\nu_\mu,\mu^-\pi^+)p$ at neutrino energy
$E_\nu = 2$ GeV and $Q^2=$0.1(solid), 0.5(dashed), 1.0(dotted),
1.5(dot-dashed) (GeV/c)$^2$.}
\label{fig:fig15}
\end{figure}

\end{document}